# Latent Signal Models: Learning Compact Representations of Signal Evolution for Improved Time-Resolved, Multi-contrast MRI

Yamin Arefeen[1], Junshen Xu[1], Molin Zhang[1], Zijing Dong[2], Fuyixue Wang[2], Jacob White[1], Berkin Bilgic[2,3], Elfar Adalsteinsson[1,4,5]

1 Department of Electrical Engineering and Computer Science, Massachusetts Institute of Technology, Cambridge, MA, USA
2 Athinoula A. Martinos Center for Biomedical Imaging, Charlestown, MA, USA
3 Department of Radiology, Harvard Medical School, Boston, MA, USA
4 Harvard-MIT Health Sciences and Technology, Massachusetts Institute of Technology, Cambridge, MA, USA
5 Institute for Medical Engineering and Science, Massachusetts Institute of Technology, Cambridge, MA, United States.

**Corresponding Author:**

Yamin Arefeen

yarefeen@mit.edu

Body word count:   ~5000

Body Figure Count: 9

# ABSTRACT

**Purpose:** Training auto-encoders on simulated signal evolution and inserting the decoder into the forward model improves reconstructions through more compact, Bloch-equation-based representations of signal in comparison to linear subspaces.

**Methods:** Building on model-based nonlinear and linear subspace techniques that enable reconstruction of signal dynamics, we train auto-encoders on dictionaries of simulated signal evolution to learn more compact, non-linear, latent representations. The proposed Latent Signal Model framework inserts the decoder portion of the auto-encoder into the forward model and directly reconstructs the latent representation. Latent Signal Models essentially serve as a proxy for fast and feasible differentiation through the Bloch-equations used to simulate signal. This work performs experiments in the context of $T_2$-shuffling, gradient echo EPTI, and MPRAGE-shuffling. We compare how efficiently auto-encoders represent signal evolution in comparison to linear subspaces. Simulation and in-vivo experiments then evaluate if reducing degrees of freedom by incorporating our proxy for the Bloch-equations, the decoder portion of the auto-encoder, into the forward model improves reconstructions in comparison to subspace constraints.

**Results:** An auto-encoder with one real latent variable represents FSE, EPTI, and MPRAGE signal evolution as well as linear subspaces characterized by four basis vectors. In simulated/in-vivo $T_2$-shuffling and in-vivo EPTI experiments, the proposed framework achieves consistent quantitative NRMSE and qualitative improvement over linear approaches. From qualitative evaluation, the proposed approach yields images with reduced blurring and noise amplification in MPRAGE shuffling experiments.

**Conclusion:** Directly solving for non-linear latent representations of signal evolution improves time-resolved MRI reconstructions through reduced degrees of freedom.

# 1    INTRODUCTION

Efficient time-resolved Magnetic Resonance Imaging (MRI) enables a wide range of applications such as MRI spectroscopy[1], quantitative parameter mapping[2,3], motion-resolved imaging[4,5], and blur-free, multi-contrast imaging from fast-spin-echo[6] (FSE) and Magnetization Prepared Rapid Acquisition Gradient Echo[7,8] (MPRAGE) acquisitions.  In this manuscript, we focus on applications that reconstruct individual echo-images with differing contrasts from acquisitions with evolving signal evolution over an echo-train[7].

Time-resolving MRI requires lengthy scan times using traditional methods.  Compressed sensing[9,10] and parallel imaging[11,12] reduce acquisition times in structural MRI, but these techniques in isolation do not enable clinically feasible time-resolved MRI.  Machine learning provides another avenue for acceleration.  Models, typically neural networks trained in a supervised fashion, impose custom regularization or directly reconstruct undersampled data[13–17].  Some work demonstrates applications of supervised and unsupervised machine learning in dynamic imaging[18,19], but the cost of acquiring fully sampled data hampers widespread use of supervised learning in many time-resolved acquisitions.

Time-resolved MRI employs a variety of different sequence and reconstruction techniques to reduce acquisition times in the presence of the additional imaging dimension.  Some techniques model signals with an analytic formula and then resolve the underlying tissue parameters that characterize this model[20–25], introducing a nonlinear optimization problem that may not account for slice-profile effects, stimulated-echoes, and B1+ inhomogeneity[26].  Other techniques build the Bloch-equations into the forward model and solve for the underlying parameters governing the model[27–29], but require extensive computation or approximation of gradients.

Alternatively, linear subspace-based constraints exploit temporal redundancies to significantly reduce the degrees of freedom in the reconstruction problem[3,6,30–34].  Low-rank constraints[3,35,36] combined with compressed sensing[9,10] and parallel imaging[11,12] also synergistically improve capability of resolving signal evolution.

For example, the highly-under-sampled shuffling[6,7,37,38] and echo-planar-time-resolved-imaging[2,39] (EPTI) techniques resolve multi-contrast images from fast-spin-echo[40] (FSE, TSE), MPRAGE, and EPI acquisitions. These techniques simulate dictionaries of signal evolution and generate linear subspaces using the singular-value-decomposition (SVD). The forward models explicitly incorporate subspace constraints and directly solve for (and impose regularization[35] on) the associated unknown linear coefficients with significantly reduced degrees of freedom. Shuffling reconstructs multi-contrast images with improved sharpness as signal evolutions no longer modulate k-space. Similarly, EPTI produces distortion- and blurring free images across the EPI readout. Shuffling yields positive clinical outcomes[38], while EPTI enables distortion/blurring free, rapid quantitative mapping, diffusion imaging[41], and fMRI[42].

However, linearization of the nonlinear reconstruction problem may induce more degrees of freedom than what exists in the underlying physical process. Additionally, linearization requires solving for both the real and imaginary portion of each complex linear coefficient, further increasing the number of unknowns.

Recently, auto-encoders trained on dictionaries of simulated signal in MRI spectroscopy and diffusion acquisitions learn compact latent representations of signal space[43,44]. The proposed MRSI and diffusion techniques employ the trained auto-encoders as regularization in reconstruction.

We propose Latent Signal Models, combining ideas from Bloch-equation models, linear subspaces, and latent representations for improved time-resolved MRI reconstruction. Latent Signal Models simulate dictionaries of signal evolution to train an auto-encoder to learn a compact representation of signal. The proposed reconstruction framework then incorporates the decoder portion, of the trained auto-encoder, into the imaging forward model and solves for the learned latent representation of signal. This improves reconstruction quality with reduced degrees of freedom in comparison to linear techniques. The decoder produces the time-series of multi-contrast images from the reconstructed latent representation. Latent Signal Models learn to represent the Bloch-equations with a simple neural-network that can be inserted into the forward model enabling reconstruction with the sophisticated auto-differentiation tools developed for machine learning[45]. In this way, we feasibly, quickly, and conveniently solve a reconstruction problem that maintains the benefits of and circumvents the demanding optimization usually induced by incorporating the Bloch-equations into the forward model.

We begin by introducing and characterizing our Latent Signal Model framework through application in $T_2$-shuffling. We show that auto-encoders learn more compact representations of FSE signal evolution in comparison to linear models. Then, simulation and retrospectively under-sampled in-vivo reconstruction experiments suggest that the reduced degrees of freedom afforded by the Latent Signal Model framework improves reconstruction accuracy in comparison to linear subspace constraints. Additional experiments analyze the stability of the technique across varying noise instances. Next, we empirically verify that the trained auto-encoder essentially serves as a fast and feasible proxy for optimization through the simulation of signal evolution. Finally, we demonstrate versatility through improved reconstructions in in-vivo, retrospectively under-sampled gradient-echo EPTI, and under-sampled MPRAGE shuffling experiments.

## 2  METHODS

### 2.1. Time-resolved MRI with Linear Subspace Constraints

Sequences that acquire multiple k-space lines after excitation or inversion across a gradient-echo or spin-echo train concatenate data from all echoes into a single k-space matrix, often of dimension $M \times N \times P \times C$, where $M, N, P$ represent read-out, phase-encode, and partition dimensions respectively and $C$ represents the number of coils. Parallel imaging then produces a single image which suffers from blurring[46] or distortion[47] artifacts due to signal decay and phase evolution during the echo train.

To improve image sharpness and reduce distortion, techniques like EPTI and shuffling aim to resolve $x \in C^{M \times N \times P \times T}$, a time-series of multi-contrast images, where $T$ represents the number of echoes or timepoints, by considering the temporal representation of acquired k-space, $y \in C^{M \times N \times P \times C \times T}$, which assigns k-space points to their associated time of acquisition[2,6]. However, the resultant reconstruction problems become heavily under-determined as resolving the time dimension increases the number of unknowns by a factor of $T$.

To produce tractable reconstruction problems, linear subspace techniques generate a dictionary, $D$, of simulated signal evolution from realistic tissue relaxation parameters for the desired sequence of interest. Then the first $B$ singular vectors of $D$ form a low dimensional subspace $\Phi \in C^{T \times B}$ that can be inserted into the forward model to produce:

$$\alpha^* = argmin_\alpha ||y - F\,S\,H\,\Phi\,\alpha|| + \lambda\,R(\alpha) \qquad \text{Eq1}$$

Where $F$ and $S$ represent the under-sampled Fourier and Coil-sensitivity operators applied to each time-point, $H$ represents temporally varying phase (if necessary), $\alpha \in C^{M \times N \times P \times B}$ represents the subspace coefficients the optimization problem solves for, and $R$ represents a spatial-regularization function applied to the subspace coefficients. The estimated coefficients $\alpha^*$ yields the time-series of images with $x^* = \Phi\alpha^*$. This formulation reduces the number of unknowns by $\frac{T}{B}$. For example, T$_2$-shuffling reconstructs data from an FSE sequence with $T = 80$ and utilizes subspaces with $B = 4$, reducing unknowns by a factor of 20×.

## 2.1. Latent Signal Models Reconstruction Framework for Time-resolved MRI

We propose learning a nonlinear latent representation of the signal evolution dictionary, $D$. Let $E_\Theta$ and $Q_\Psi$ represent fully-connected neural-networks for the encoder and decoder of an auto-encoder[48,49]. The auto-encoder learns a latent representation of signal evolution by minimizing the following with respect to its weights and biases, $\Theta$ and $\Psi$:

$$\Theta^*, \Psi^* = argmin_{\Theta,\Psi} ||D - Q_\Psi(E_\Theta(D))||_2^2 \qquad \text{Eq2}$$

The trained decoder can then be inserted in the forward model,

$$\beta^*, \rho^* = argmin_{\beta,\rho} \left|\left|y - F\,S\,H\,[\rho Q_{\Psi^*}(\beta)]\right|\right|_2^2 + \lambda\,R(\beta,\rho) \qquad \text{Eq3}$$

to reconstruct the latent representation $\beta^* \in R^{M \times N \times P \times L}$, proton density $\rho^* \in C^{M \times N \times P}$, and time-series of images with $x^* = \rho^* Q_{\Psi^*}(\beta^*)$, where $L$ is the number of latent variables. Unlike previous work[43,44], the decoder prior does not require tuning an additional regularization parameter, and separate spatial regularization can be applied directly to the unknown latent representation and density. **Figure 1** visualizes the Latent Signal Model framework in comparison to the standard linear approach through an exemplar T$_2$-shuffling setting.

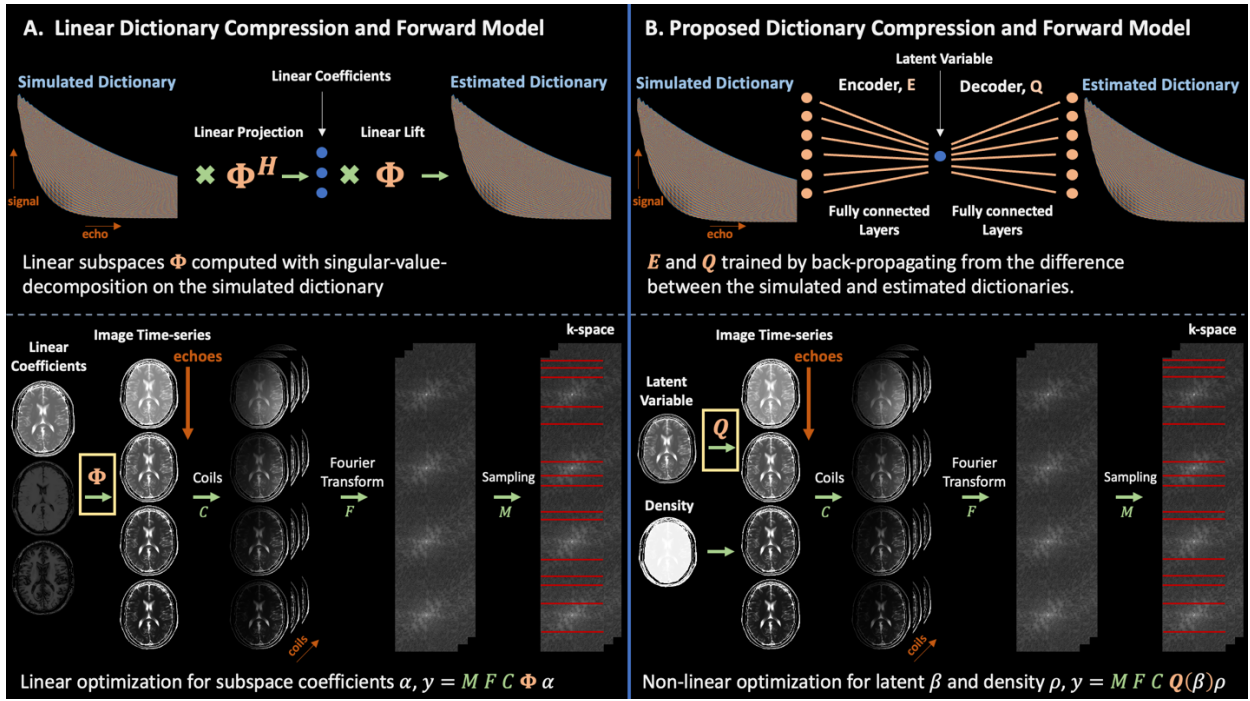

**Figure 1:** Comparing linear subspace constraints and the proposed Latent Signal Model framework in the $T_2$-shuffling setting. To represent FSE signal evolution, (A) the SVD generates a subspace from a dictionary of signal evolution with 2-4 coefficients, while (B) the auto-encoder learns a more compact representation of signal evolution with 1 latent variable. Inserting either the subspace or decoder in the forward model reduces the number of variables required to resolve signal dynamics. The auto-encoder's more compact latent representation, compared to linear subspaces', enables improved reconstruction quality in subsequent experiments.

With complex-valued coefficients $\alpha$, linear reconstructions resolve $M \times N \times P \times B \times 2$ unknowns; while the real-valued latent representation $\beta$ and the complex-valued proton density $\rho$ yield $M \times N \times P \times (L + 2)$ unknowns in the proposed framework. If $L + 2 < 2 \times B$, the Latent Signal Model framework produces a reconstruction problem with fewer unknowns. In subsequent experiments, linear models require $B = 2, 3$, while our method uses $L = 1$, corresponding to a 1.5× to 2.5× reduction of unknowns.

In essence, the decoder neural-network serves as a proxy for the simulation process used to generate the dictionary of signal evolution and synergizes with auto-differentiation tools for fast and feasible optimization.
88

## 3.1. T$_2$-shuffling Experiments

Experiments begin in T$_2$-shuffling[6] settings for demonstration and characterization, while subsequent results showcase the technique's versatility through gradient-echo EPTI and MPRAGE-shuffling.

### 3.1.1. Dictionary Compression

We compare how efficiently SVD-generated linear-subspaces and a trained auto-encoder represent FSE signal evolution. Extend-phase-graph (EPG) simulations[50] generated training and testing dictionaries of signal-evolution for T$_2$ = 50-400 ms, T$_1$ = 1000 ms, 80 echoes ($T = 80$), 160 degree refocusing pulses, and 5.56 ms echo spacing. From the training dataset, SVD generated linear subspaces with {1,2,3,4} singular vectors, while minimizing Eq3 produced auto-encoders with {1,2,3,4} latent variables.

Each auto-encoder utilized two fully-connected layers for both the decoder and encoder with Hyperbolic Tangent[51] (tanh) activation functions. $T \times 1$ vectors representing signal evolutions served as inputs to the auto-encoder. The first layer of the encoder transforms the $T \times 1$ input vector to $\frac{T}{2} \times 1$, and the second layer takes the $\frac{T}{2} \times 1$ vector to the latent $L \times 1$ representation. The decoder performs the same operations, characterized by different weights, in reverse. Linear subspaces and auto-encoders compressed and reconstructed signal in the testing dictionary, and the resultant reconstruction accuracies were compared.

### 3.1.2. Simulated Reconstruction Experiments

Next, we simulated T$_2$-shuffling acquisitions from a numerical phantom with T$_1$, T$_2$, proton density, and 8-channel coil-sensitivity maps[52]. EPG simulations produced a time-series of k-spaces (with added noise) for each echo of a FSE sequence: $T = 80$ echoes, 5.56 ms echo spacing, and 160 degree refocusing pulses. We applied an undersampling mask that mimics a 2D T2-shuffling acquisition with 4-shots that samples a random phase-encode line at each echo and labeled the dataset *simulated_t2shfl1*. This corresponds to 320 sampled k-space lines, resulting in an overall acceleration factor with respect to the desired time-series of images to reconstruct of $R = \frac{256 * 80}{320} = 64$.

The experiment compares linear subspace reconstructions using {4,6} degrees of freedom (or {2,3} complex coefficients) and wavelet or locally-low-rank regularization, to the proposed Latent Signal Model reconstruction with 3 degrees of freedom and wavelet regularization (1 latent variable, 1 real proton density, 1 imaginary proton density). The regularization parameter, $\lambda$, in all reconstructions was experimentally optimized to minimize error. Auto-differentiation in PyTorch[53] with a GPU compatible implementation of Wavelets[54] solved the proposed reconstruction, while the BART[55] toolbox reconstructed the linear problem.

### 3.1.3. Simulated Auto-encoder Hyperparameter Ablation Experiments

With identical sequence parameters, we generated an additional dataset, labeled *simulated_t2shfl2*, modeling a 3D T2-shuffling acquisition by assuming each echo-k-space corresponds to phase and partition encode dimensions and sampling $256 \times 256$ k-space points randomly throughout the 80 echoes[6]. An ablation study trained auto-encoders on all combinations of the following hyperparameters and compared the resultant un-regularized Latent Signal Model reconstructions. Since these experiments do not employ regularization, we utilized a dataset with more incoherence to better condition the reconstruction problem.

- Nonlinearity: [relu, tanh, leakyrelu]
- Training Epochs: [20K, 100K]
- Learning Rate: [1e-4, 1e-5]
- FC Layers: [1,2,3]
- Latent variables: [1]

### 3.1.4. Retrospective in-vivo reconstruction experiments

All imaging protocols were performed with approval from the local institutional review board with written informed consent. A volunteer was scanned with a Siemens 3T Trio (Siemens Healthineers, Erlangen, Germany) system using a 12-channel receive head coil. A fully-sampled spatial and temporal multi-echo T$_2$-weighted dataset with field-of-view = 180 mm x 240 mm, matrix size = 208 x 256 matrix size, slice thickness = 3 mm, echo spacing = 11.5 ms, and $T = 32$ echoes was acquired.

The applied undersampling mask generated acquisitions with 7 shots that samples a random phase-encode line at each echo, and we label the dataset *invivo_t2shfl1*. This corresponds to 224 sampled k-space lines, resulting in an overall acceleration of $R = \frac{256 * 32}{224} \approx 36$.

The experiment compares linear subspace reconstructions with {4,6} degrees of freedom and wavelet or locally-low-rank regularization to the proposed Latent Signal Model reconstruction with 3 degrees of freedom and wavelet regularization. Additionally, the proposed approach utilized an auto-encoder with 3 fully-connected layers and tanh activation functions. We chose all experimentally optimized regularization parameters to minimize error, and the auto-differentiation and BART frameworks solved the proposed and linear problems respectively.

### 3.1.5. Retrospective Auto-encoder Hyperparameter Ablation Experiments

On a similar 8-shot dataset, labeled *invivo_t2shfl2*, an ablation study trained auto-encoders on combinations of the following hyper-parameters and compared the resultant un-regularized Latent Signal Model reconstructions. Since these experiments do not employ regularization, we utilized a dataset with more shots to better condition the reconstruction problem.

- Nonlinearity:        [tanh, leakyrelu]
- Training Epochs:     [20K, 100K, 200K]
- Learning Rate:       [1e-4, 1e-5]
- FC Layers:           [2,3]
- Latent variables:    [1]

### 3.1.6. Analyzing the effects of reduced degrees of freedom and evaluating reconstruction stability across various noise instances

The aforementioned experiments combined temporal constraints, spatial regularization, and tuned hyper-parameters for best reconstruction performance. Thus, the following experiment isolates and characterizes the improvements afforded by the reduced degrees of freedom from Latent Signal Models and evaluates stability of the reconstruction across various noise instances in the spirit of g-factor[12] analysis.

We added 250 instances of realistic gaussian noise to the *simulated_t2shfl2* dataset, described in the simulation hyper-parameter experiment, and 2D-, 7-shot, in-vivo *invivo_t2shfl1* dataset. Then, the proposed Latent Signal Model framework with 3 and linear subspaces with {4,6} degrees of freedom reconstructed the 250 k-spaces in both the simulation and retrospective regimes without regularization. We report the mean and standard deviation of the normalized-root-mean-squared-error (NRMSE) at each reconstructed echo time and display mean reconstructed error maps.

### 3.1.7. Latent Signal Models Learn a Proxy to Feasibly and Efficiently Differentiate through MRI Signal Simulations in a Reconstruction Setting

The subsequent experiment demonstrates that the trained auto-encoder serves as a proxy for fast and feasible optimization through the signal simulation process directly in the reconstruction problem.

Let $G$ represent the simulation function that produces signal evolution given an input $T_2$ and proton density in $T_2$-shuffling. The simulation-based reconstruction problem becomes:

$$argmin_{T_2,\rho} V(T_2, \rho) = ||y - F\,S\,H\,G(T_2, \rho)||_2^2 \qquad \text{Eq4}$$

which solves for the $T_2 \in R^{M \times N}$ map and density $\rho \in C^{M \times N}$.

We show that minimizing Eq3 in the Latent Signal Model framework produces a solution that also minimizes the simulation-based reconstruction problem in Eq4. We use the proposed framework to reconstruct the *simulated_t2shfl2,* and *invivo_t2shfl2* datasets (both without regularization). The optimization runs for 1000 iterations, applying the following procedure every 50 iterations:

- Extract the current latent variable and density estimates, $\beta^i$ and $\rho^i$.
- With $\beta^i$, estimate a T$_2$-map, $T_2^i$, using dictionary matching directly in latent variable space[56].
- Compute and record the L$_2$-norms of the following gradients $||\nabla_{T_2}(V(T_2^i, \rho^i))||_2$ and $||\nabla_\rho(V(T_2^i, \rho^i))||_2$.

Then solving Eq3 also finds a (local) minima of Eq4 if $||\nabla_{T_2}(V(T_2^i, \rho^i))||_2$ and $||\nabla_\rho(V(T_2^i, \rho^i))||_2$ approach 0 as the proposed framework's iteration count, $i$, increases. We also take the final solution from the proposed framework, $\beta^*$ and $\rho^*$, estimate a T$_2$-map, $T_2^*$, and solve the simulation-based reconstruction problem in Eq4 using $T_2^*$, and $\rho^*$ as initialization. If the proposed framework appropriately minimizes Eq4, then the solution will not change significantly from the initial guess.

Finally, we compare computation times in optimizing the Latent Signal Model and simulation-based frameworks, and we try applying the simulation based forward model (Eq4) to reconstruct the simulated T$_2$-shuffling dataset without using the proposed framework's solution as initialization.

We implement $G$ with a PyTorch EPG algorithm[57] that enables computation of gradients with auto-differentiation.

### 3.2. Gradient-Echo EPTI Experiments

Here, we demonstrate application of Latent Signal Models in 2D-gradient-echo (GE) EPTI[2,58] which continuously measures k-space during T$_2$* dominated signal decay after an initial 90-degree excitation pulse. GE-EPTI resolves signal dynamics from the highly-undersampled k$_x$-k$_y$-t dataset through B$_0$-informed linear subspace reconstructions.

A fully-sampled spatial and temporal multi-echo gradient-echo EPTI dataset with matrix size = 216 x 216, slice thickness = 3mm, 1.1 x 1.1 mm in-plane resolution, echo spacing = .93 ms, $T\ =$

*40* echoes, and 32 coils was acquired. To produce a challenging case, the under-sampling mask modeled a GE-EPTI acquisition with two shots. This corresponds to 80 sampled k-space lines, resulting in an overall acceleration factor of $R = \frac{216 * 40}{80} = 108$.

The first experiment generated $H$ from the fully-sampled dataset. Different from $T_2$-shuffling and MPRAGE, EPTI models temporally varying phase, so both the linear and Latent Signal Model reconstructions might be more sensitive to bias in $B_0$ phase estimation in this highly under-sampled experiment. We started with more accurate phase estimates to evaluate reconstructions independent of the $B_0$-estimation algorithm.

We then repeated the experiment using a $B_0$-map estimated from the central 49 k-space lines of the first 6 echoes to characterize performance with low-resolution phase. These central k-space lines were treated as a calibration pre-scan and not amongst the 80 lines used for reconstruction.

The experiments compare linear subspace reconstructions with {4,6} degrees of freedom and locally-low-rank regularization to the proposed reconstruction with 3 degrees of freedom (1 latent variable) and wavelet regularization. The proposed approach trained an auto-encoder with two-fully connected layers for both the encoder and decoder with tanh activations on a dictionary of simulated GE-EPTI signal evolution with $T_2^*$ values in the range of 10 - 300 ms. We chose all experimentally optimized regularization parameters to minimize error.

### 3.3 MPRAGE Shuffling Experiments

MPRAGE-shuffling[7] employs linear subspaces to resolve multiple image contrasts across the MPRAGE echo train. We apply the proposed framework to model MPRAGE signal and reconstruct under-sampled MPRAGE-shuffling data.

### 3.3.1 MPRAGE Dictionary Compression

First, we compare how efficiently SVD-generated linear-subspaces and a trained auto-encoder represent MPRAGE signal evolution. Bloch-simulations generated training and testing dictionaries of MPRAGE signal-evolution for $T_1$ = 500-3000 ms, 256 echoes ($T = 256$), 8 degree flip angle, and 7.8 ms echo spacing. Linear subspaces with {1,2,3,4} singular vectors and auto-encoders with {1,2,3,4} latent variables were generated from the training set. Each auto-encoder utilized two fully connected layers for both the decoder and encoder with Hyperbolic Tangent (tanh) activation functions. Resultant compression accuracies in the testing dictionary were compared.

### 3.3.2 MPRAGE Reconstruction Experiments

MPRAGE shuffling data was acquired using a spatial matrix size of 256 x 256 x 256, 1 mm isotropic resolution, 32-channel head coil, turbo factor (echo-train-length) of 256, 1100 ms inversion time, 2500 ms TR, 7.8 ms echo-spacing, and 648 second total acquisition time. The 648 second acquisition time corresponds to a fully-sampled non-time-resolved MPRAGE acquisition, but the MPRAGE-shuffling sequence distributed the phase/partition encode points randomly throughout the echo-train. 16-channel SVD coil-compression[59] and a read-out fourier transform was applied to process an axial-slice for the experiment. The data were further under-sampled by excluding all k-space points measured after 324 seconds, resulting in an overall acceleration factor of $R = \frac{256*256*256}{256*128} = 512$.

The reconstruction experiment compares linear subspace constrained reconstructions with 4 degrees of freedom and wavelet or locally-low-rank regularization to the Latent Signal Model reconstruction with 3 degrees of freedom, wavelet regularization, and the MPRAGE trained auto-encoder.

All reconstructions were performed with varying regularization levels, and the best qualitative images are displayed. Auto-differentiation in PyTorch and BART solved the proposed and linear problems respectively.

## 4 RESULTS

### 4.1. T$_2$-shuffling Experiments

#### 4.1.1. Dictionary Compression

In **Figures 2** (A) and (B), the proposed auto-encoder reconstructs the training dictionary with {0.10%, 0.05%, 0.05%, 0.03%} average NRMSE and the testing dictionary with {0.11%, 0.05%, 0.06%, 0.04%} NRMSE with {1,2,3,4} latent variables. Linear subspaces with {1,2,3,4} coefficients yield {18.41%, 3.18%, 0.44%, 0.05%} and {18.68%, 3.30%, 0.47%, 0.06%} NRMSE on the training and testing dictionaries. **Figure 3** (C) shows plots of RMSE versus T$_2$-value on individual signal-evolution entries from the testing dictionary; the auto-encoder with one latent variable uniformly outperforms {1,2,3} and matches performance of 4 linear coefficients.

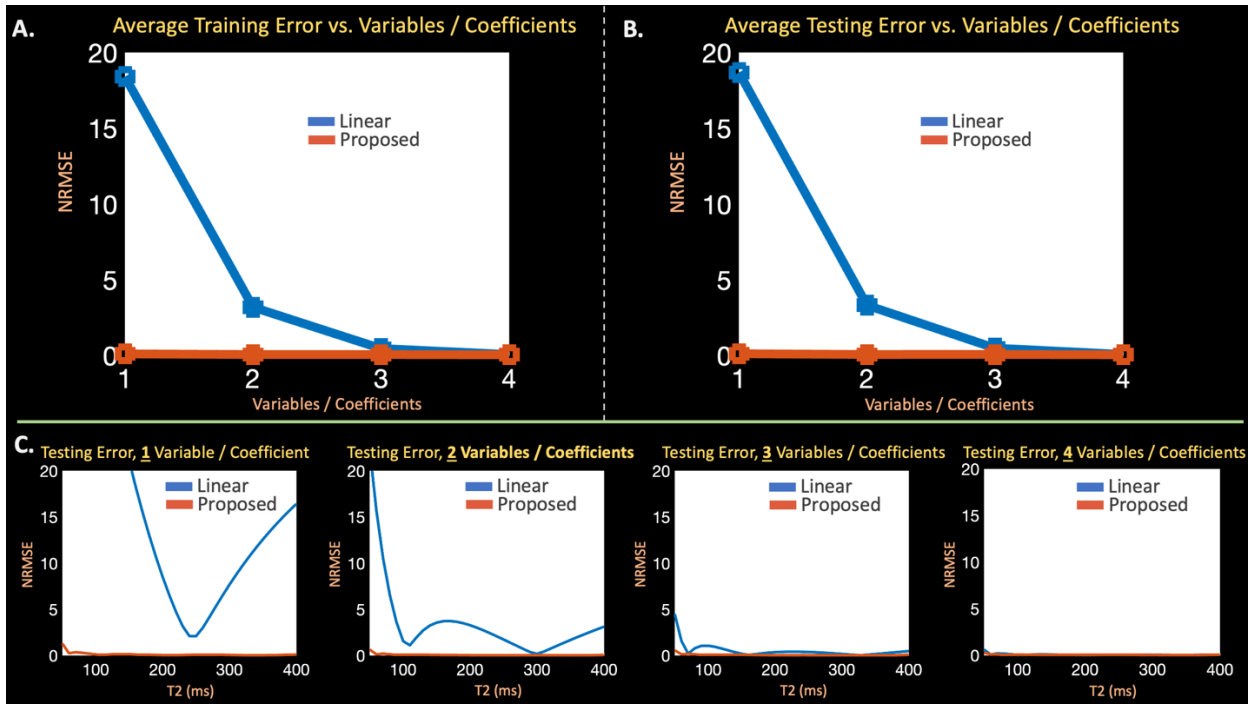

**Figure 2:** (A) + (B) Average error across the entire training and testing dictionaries of FSE signal evolution using linear subspaces and proposed auto-encoders with {1,2,3,4} coefficients or latent variables. (c) Error for individual signal evolutions, associated with different T2 values, in the testing dictionary using subspaces and auto-encoders. The subspace requires 3-4 coefficients, while the auto-encoder effectively captures signal evolution with just 1 latent variable.

#### 4.1.2. Simulated Reconstruction Experiments

Simulation experiments in **Figure 3** compare the proposed Latent Signal Model reconstruction with one latent variable (three degrees of freedom) and wavelet regularization to the linear reconstructions with {2,3} complex coefficients ({4,6,8} degrees of freedom) and wavelet or locally-low-rank regularization on the *simulated_t2shfl1* dataset. Selected echo images and error maps in (A) demonstrate that linear reconstructions with 2 coefficients cannot adequately represent signal, while 3 linear coefficients exhibit increased noise amplification. The proposed approach accurately represents signals and improves reconstruction performance. In (B), the proposed approach achieves lower NRMSE across all echoes. Linear wavelet with {4,6} degrees of freedom achieves {9.6,4.6}% average NRMSE across all echoes, while the proposed approach achieves 3.2%.

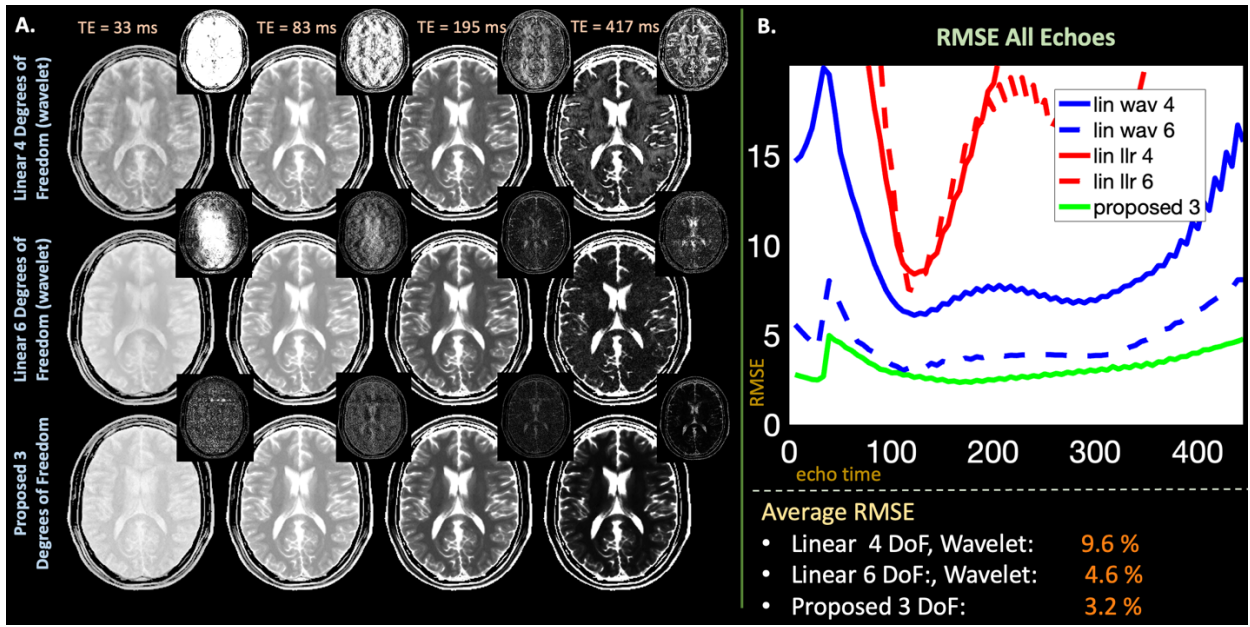

**Figure 3:** (A) Exemplar reconstructed echo-images, associated error maps, and (B) RMSE plot vs. echo time from a simulated T2-shuffling acquisition comparing the subspace and proposed Latent Signal Model reconstruction with {4,6} and 3 degrees of freedom respectively. The reduced degrees of freedom enabled by inserting the decoder into the $T_2$-shuffling forward model yields cleaner images and lower RMSE.

### 4.1.3. Simulated Auto-encoder Hyperparameter Ablation Experiments

**Supporting Figure 1** reports results from the Latent Signal Model hyper-parameter ablation experiment for the *simulated_t2shfl2* dataset. (A) displays a grid of average NRMSE across all echoes for different combinations of non-linearities and auto-encoder layers with best choice of step-size and learning rate. Hyperbolic tangent with 2 model layers achieved the best NRMSE. (C) plots the performance of hyperbolic tangent and leakyRelu trained with the same hyperparameters, but with 5 different random initialization of model parameters. LeakyRelu performance varies across the different repetitions, while hyperbolic tangent gives consistent results.

### 4.1.4. Retrospective in-vivo reconstruction experiments

In-vivo, retrospectively under-sampled experiments in **Figure 4** compare the proposed Latent Signal Model reconstruction with one latent variable (three degrees of freedom) and wavelet regularization to the standard linear reconstruction with {2,3} complex coefficients ({4,6,8} degrees of freedom) with wavelet or locally-low-rank regularization on the *invivo_t2shfl1 dataset*. (A) displays selected echo reconstructions and error maps in which the proposed framework reduces artifacts. Moreso, the proposed technique achieves lower NRMSE, as shown in (B), with average NRMSE across all echoes of 13.8% in comparison to {15.8%, 16.8%} and {16.8%, 15.2%} of the linear reconstructions.

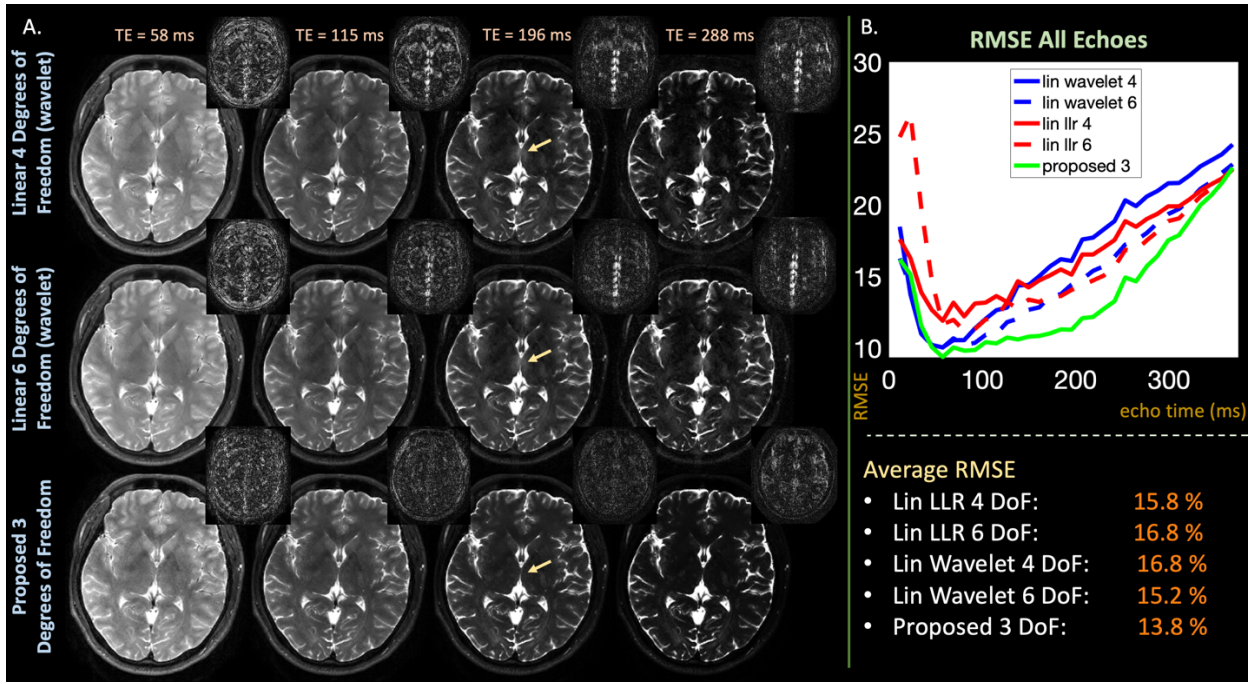

**Figure 4:** (A) Exemplar reconstructed echo images and associated error maps from the in-vivo, 7-shot retrospective, 2D-T$_2$-shuffling acquisition at exemplar echo times comparing subspace constraints with {4,6} and the proposed approach with 3 degrees of freedom. (B) RMSE plots comparing reconstructed images at all echo times. Like in simulation, the proposed approach yields cleaner images and lower RMSE through reduced degrees of freedom.

### 4.1.5. Retrospective Auto-encoder Hyperparameter Ablation Experiments

**Supporting Figure 2** illustrates hyper-parameter analysis with Latent Signal Models on the *invivo_t2shfl2 dataset*. (A) and (B) depict grids of average RMSE across all echoes for different combinations of learning rate, step-size, and number of layers using the LeakyRelu and Hyperbolic tangent nonlinearities respectively. LeakyRelu achieves lowest RMSE with 2 layers, 200K epochs, and 1e-5 learning rate, while Hyperbolic tangent achieves its minimum with 3 layers, 100K epochs, and 1e-4 learning rate. However, (C) LeakyRelu varies significantly with different parameter initializations, while tanh remains consistent.

### 4.1.6. Analyzing the effects reduced degrees of freedom and evaluating reconstruction stability across various noise instances

**Figure 5** presents results analyzing linear and proposed reconstructions without regularization across the 250 different k-space instances. In simulation (A) and in-vivo (C), the proposed approach yields lower average error maps in comparison to the linear reconstructions. Additionally, the proposed approach (B,D) achieves lower average NRMSE across all echoes, and maintains comparable variance in reconstruction accuracy.

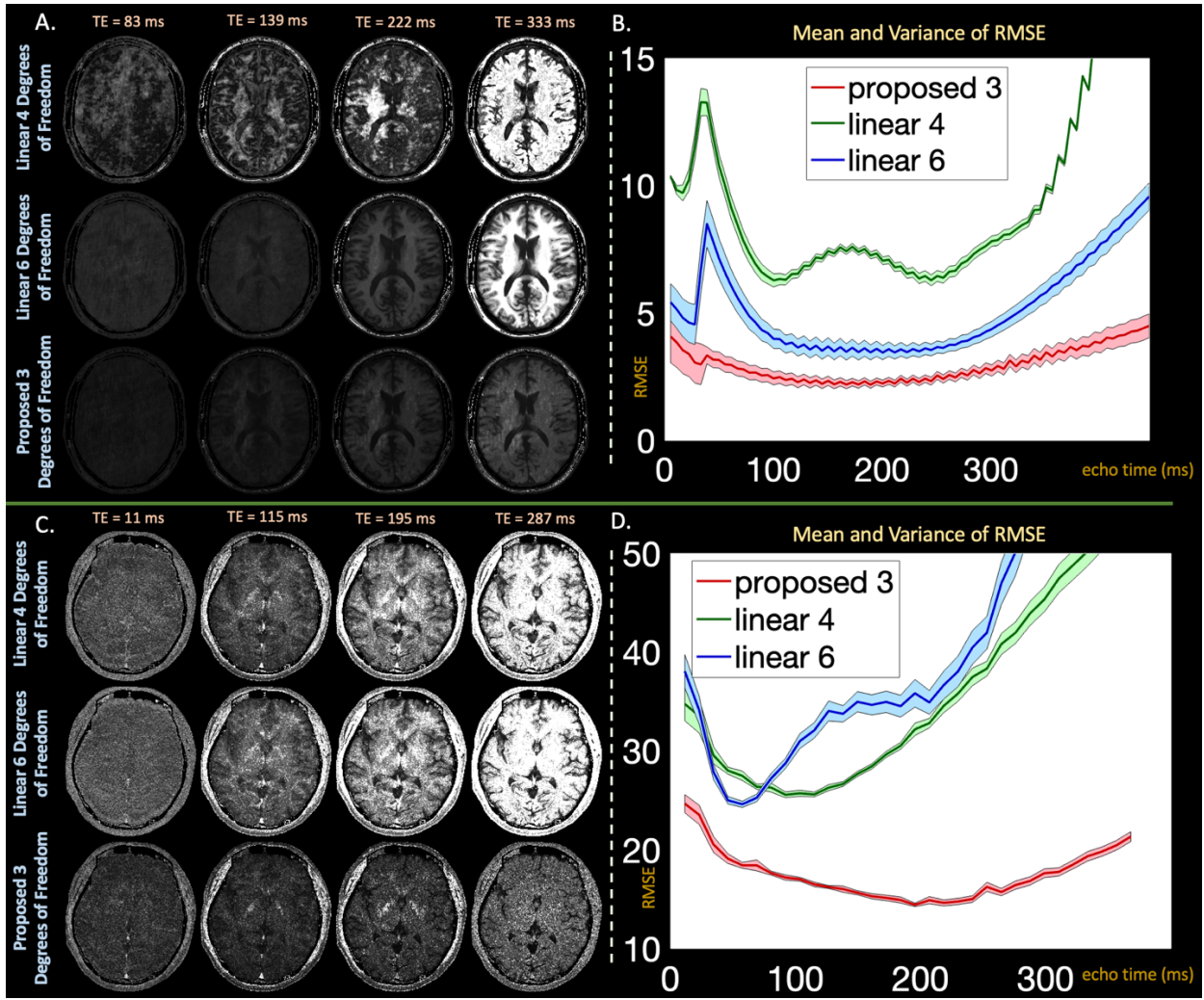

**Figure 5:** (A, C) Exemplar average error maps for the proposed and linear reconstructions across the 250 k-space instances in both the simulated and retrospective datasets. (B, D) Average and standard deviation of reconstruction RMSE at each echo. With its reduced degrees of freedom, the proposed Latent Signal Model approach achieves lower average error while maintaining similar variance across the k-space instances in-comparison to reconstructions with linear subspace constraints.

### 4.1.7. Latent Signal Models Learn a Proxy to Feasibly and Efficiently Differentiate through MRI Signal Simulations in a Reconstruction Setting

**Figure 6** (B,D) plots norms of gradients with respect to $T_2$ and density in the EPG based forward model from Eq4 as a function of every 50th Latent Signal Model optimization. In simulation,

the norm of the gradients approach 0, while the gradient norms approach a relatively small value in the retrospective experiment. **Figure 6** (A,C) displays the proposed reconstruction and the EPG-based forward model reconstruction when initialized with $T_2$ and density estimates from the proposed reconstruction. Both reconstructions look qualitatively similar, further reinforcing that the proposed optimization framework finds a solution for the EPG-based forward model.

The EPG-based forward model took 2.5 and .47 seconds per reconstruction iteration while the Latent Signal Model approach took .065 and .035 seconds on the simulated and in-vivo datasets respectively (~40x and ~13x speedups). In addition, the EPG-based forward model reconstruction on the simulated k-space did not converge to a good solution when not initialized with the proposed framework's solution.

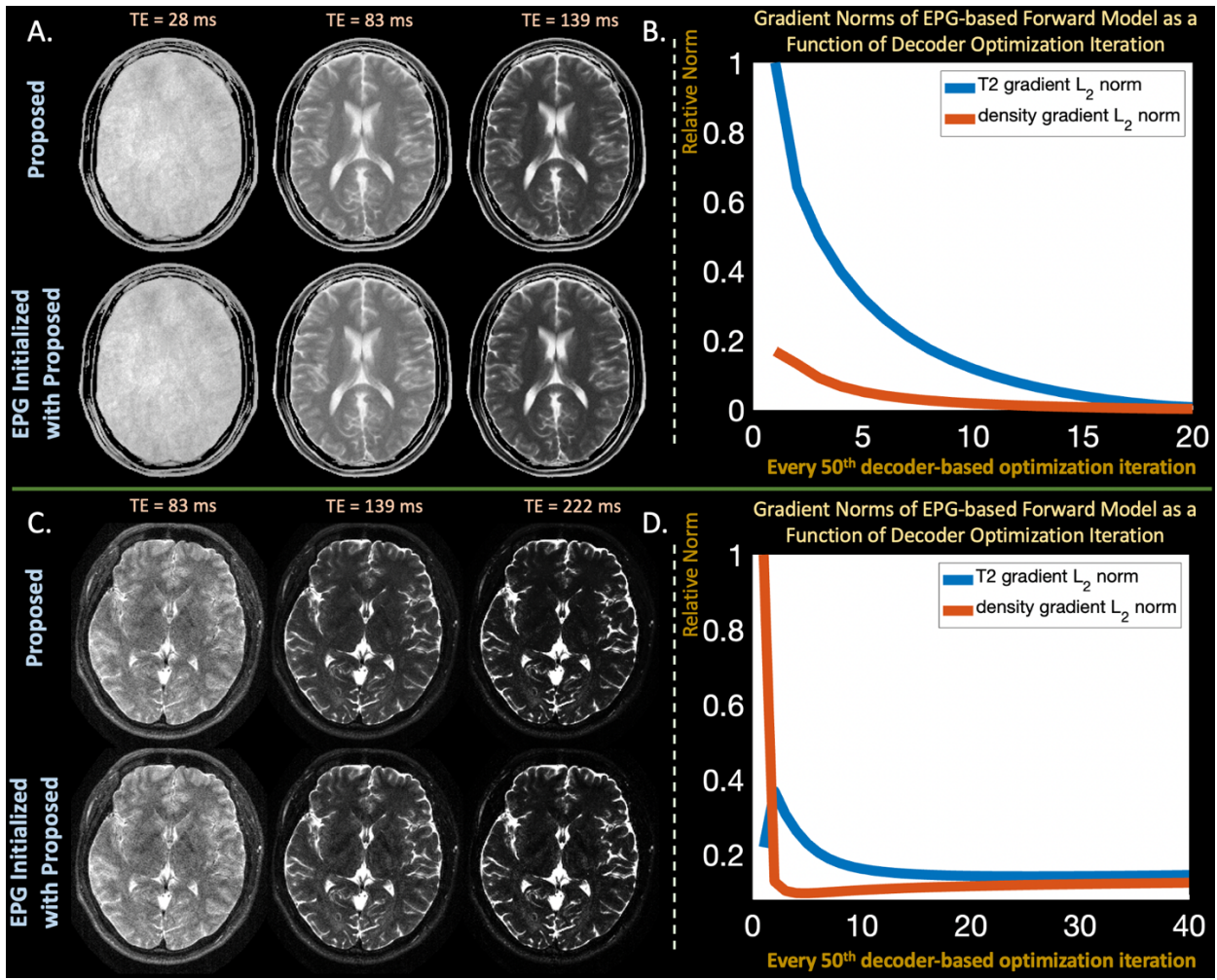

**Figure 6:** (B, D) Gradient norms with respect to $T_2$ and density of the simulation based forward model as a function of the Latent Signal Model optimization iteration. (A, C) exemplar reconstructed echo images from the proposed approach and the EPG-based forward model initialized with the proposed approach. The gradient norms approaching zero and the similarity in the reconstructions suggest that the proposed approach effectively and efficiently finds a solution to the EPG-based forward model.

### 4.2. Gradient-Echo EPTI Experiments

**Figure 7** (A) compares exemplar reconstructed echo images and error maps using linear subspaces with locally low-rank regularization and the Latent Signal Models with wavelet regularization from the in-vivo GE-EPTI dataset with phase estimated from fully-sampled data.

In this challenging case with high under-sampling, the linear reconstructions suffer from bias and artifacts, while the proposed approach significantly improves reconstruction quality. In (B), the proposed approach yields quantitative improvements, with average NRMSE across all echoes of 11.48%, in comparison to the linear average errors of {38.0%, 40.3%}.

**Supporting Figure 3** displays the same reconstructions from **Figure 7** with phase estimates calibrated from low-resolution data. Linear reconstructions remain poor, while the proposed approach still achieves significant relative improvement.

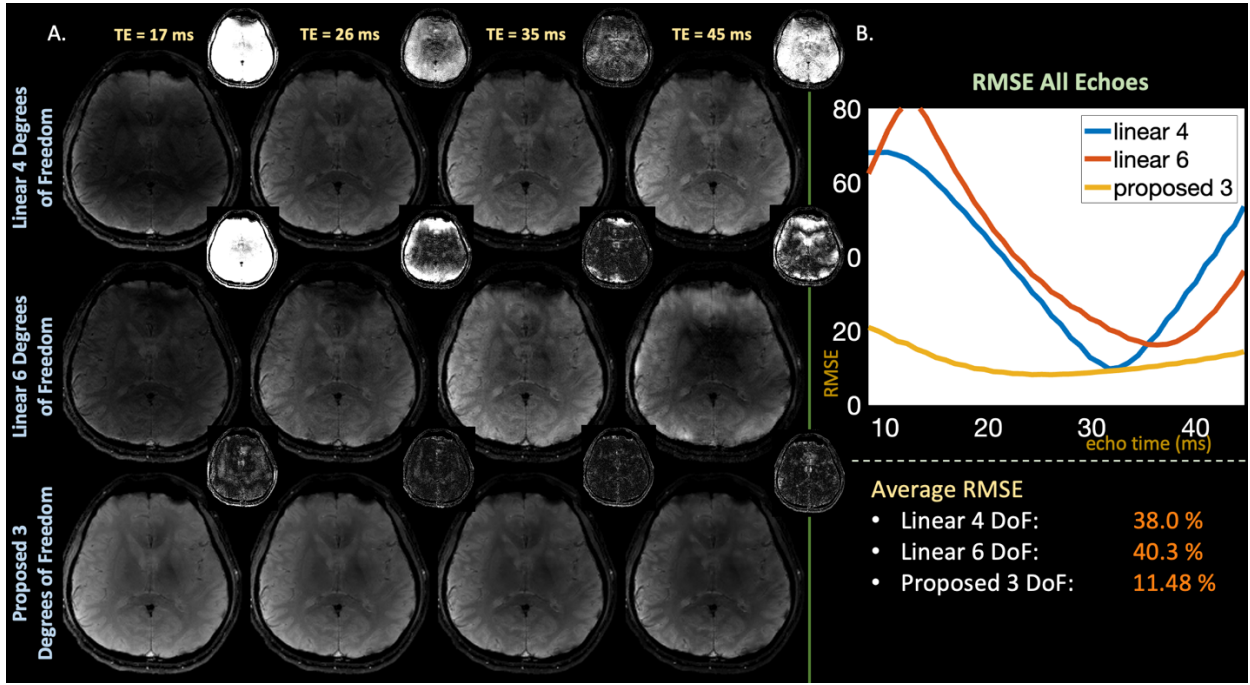

**Figure 7:** (A) Exemplar reconstructed echo images and associated error maps from the GE-EPTI dataset using phase estimated from the fully-sampled dataset, and (B) plots comparing NRMSE at each individual echo image. The proposed Latent Signal Model approach with reduced degrees of freedom significantly improves qualitative image quality and quantitative NRMSE in comparison to the linear subspace constrained reconstructions.

### 4.3. MPRAGE Shuffling Experiments

### 4.3.1 MPRAGE Dictionary Compression

In **Figures 8** (A) and (B), the proposed auto-encoder reconstructs the training MPRAGE dictionary with {0.15%, 0.14%, 0.14%, 0.13%} average NRMSE and the testing dictionary with {0.15%, 0.14%, 0.14%, 0.13%} NRMSE with {1,2,3,4} latent variables respectively. Linear subspaces with {1,2,3,4} coefficients yield {14.5%, .36%, 0.03%, 0.00%} and {15.0%, .36%, 0.03%, 0.00%} NRMSE on the training and testing dictionaries respectively. Figure 3 (C) shows plots of NRMSE versus T1-value on individual signal-evolution entries from the testing dictionary. The proposed auto-encoder with 1 latent variable represents signal evolution as effectively as the linear subspaces with 3 coefficients.

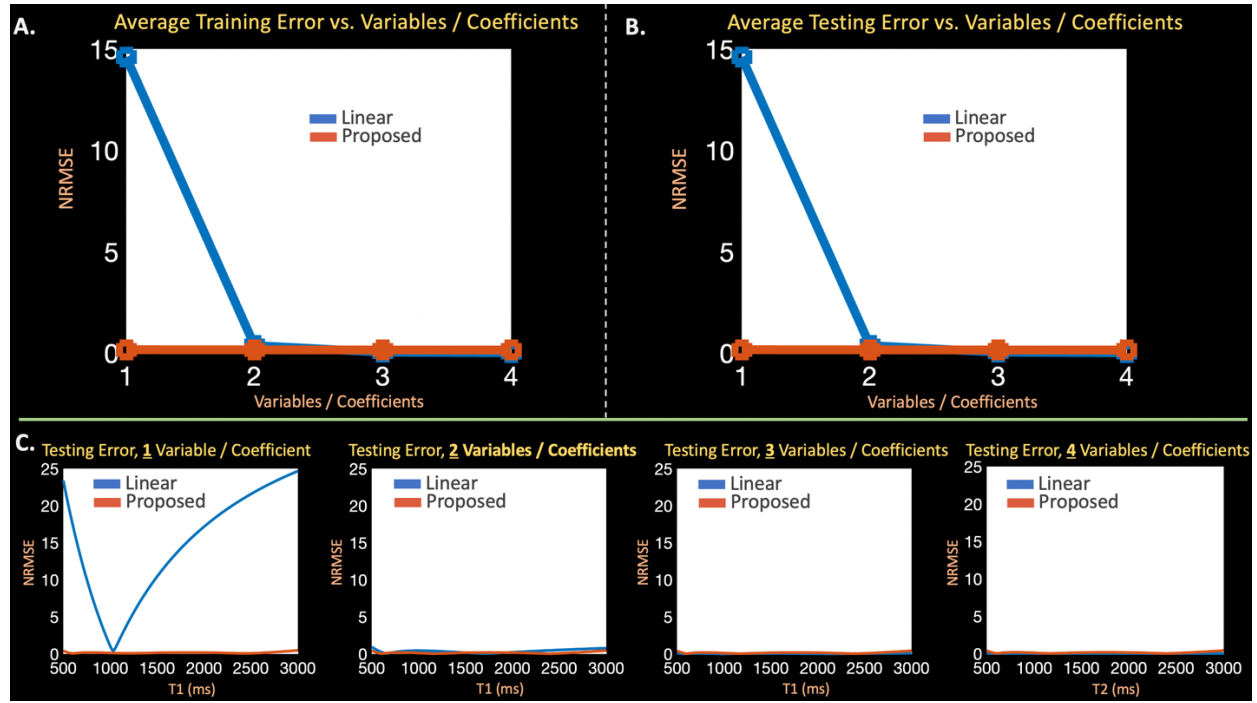

**Figure 8:** (A) + (B) Average error across the entire training and testing dictionaries of MPRAGE signal evolution using linear subspaces and proposed auto-encoders with {1,2,3,4} coefficients or latent variables. (C) Error for individual signal evolutions, associated with different $T_1$ values, in the testing dictionary using subspaces and auto-encoders. The subspace requires 3-4 coefficients, while the auto-encoder effectively captures signal evolution with just 1 latent variable.

### 4.3.2 MPRAGE Reconstruction Experiments

**Figure 9** displays reconstructions of the MPRAGE-shuffling dataset. The first two columns compare un-regularized reconstructions from the proposed and linear approaches, with 3 and 4 degrees of freedom respectively. The proposed Latent Signal Model approach significantly reduces apparent noise amplification at the displayed inversion times (3 out of 256). The third, fourth, and fifth columns display proposed with wavelet regularization and linear with wavelet or locally-low-rank regularization respectively. Particularly at TI ≈ *500* ms, the proposed approach reduces reconstruction artifacts in comparison to the linear techniques. The zoomed views suggest that the wavelet-regularized linear reconstruction reduces noise amplification at the cost of significant blurring in the image, while the locally-low-rank linear reconstruction suffers from noise amplification. The regularized Latent Signal Model reconstruction reduces artifacts and noise while maintaining sharpness.

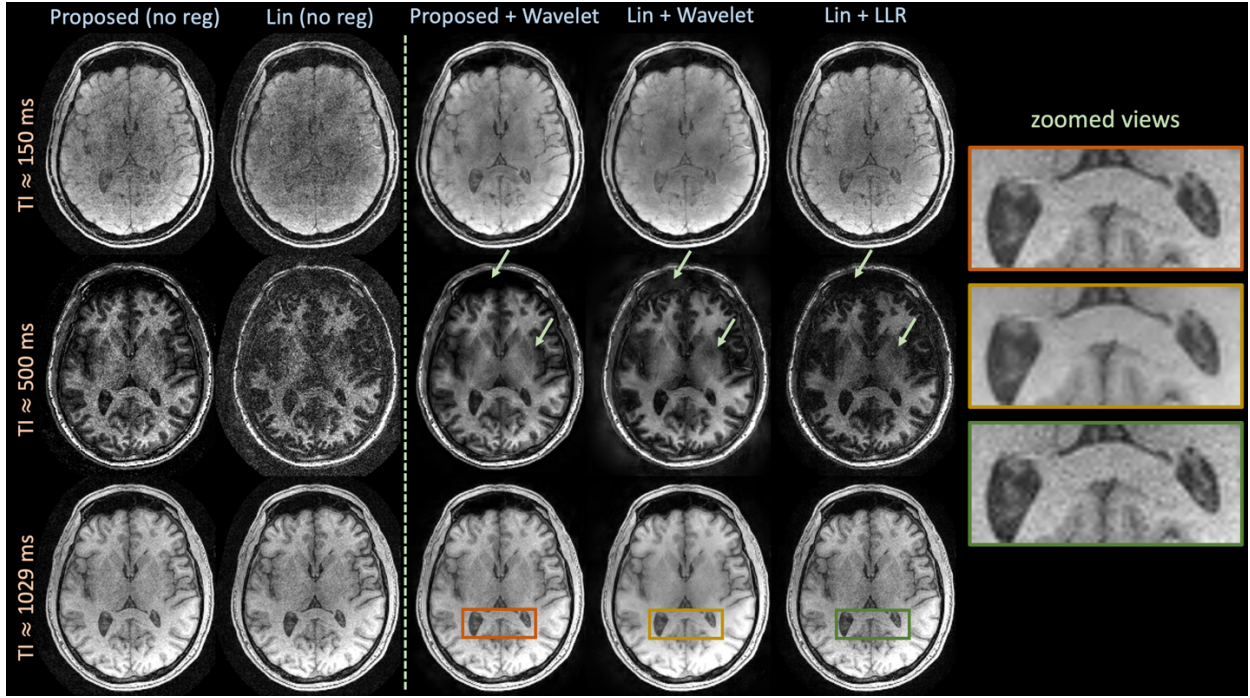

**Figure 9:** Reconstructions with and without regularization on the MPRAGE shuffling dataset comparing the proposed Latent Signal Model framework and linear subspace constraints. Without regularization, the proposed approach significantly reduces noise amplification and improves reconstruction quality. With applied regularization, linear + wavelet suffers increased blurring while linear + locally-low-rank still exhibits noise amplification. The proposed approach reduces reconstruction artifacts, particularly at TI ≈ 500 ms, improves image sharpness in comparison to linear + wavelet, and reduces noise amplification in comparison to linear + locally-low-rank.

## 5  DISCUSSION

Latent Signal Models improve time-resolved, multi-contrast MRI reconstruction by combining ideas from Bloch-equation based techniques, linear subspace constraints, and latent representation of signals. By training an auto-encoder on dictionaries of signal evolution, our technique learns more compact, latent representations of signal evolution, in comparison to linear techniques. Then, inserting the decoder into the forward model reduces degrees of freedom and serves as a proxy for fast and feasible optimization through the Bloch-equations. Our proposed method builds upon previous work that utilizes auto-encoders to regularize diffusion[43,44] and MRSI[43,44] by extending applications to multi-contrast imaging and incorporating the learned prior directly in the forward model, obviating the separate regularization term. Latent Signal Models represent FSE, EPTI, and MPRAGE signal evolution with one latent variable just as well as 3-4 SVD based linear coefficients and improve reconstruction quality in simulated and in-vivo $T_2$-shuffling, in-vivo GE-EPTI, and in-vivo MPRAGE shuffling acquisitions in comparison to linear subspace constrained reconstructions.

In the settings explored, signal largely depends on a single underlying tissue parameter ($T_2$ for FSE, $T_2^*$ for GE-EPTI, $T_1$ for MPRAGE shuffling), so we suspect that the auto-encoder successfully compresses dictionaries to just one latent variable in these applications for this reason. However, our technique does not explicitly build this knowledge into the learning algorithm; rather, the model implicitly learns to compress signal into one latent variable from the dictionary of simulated signal evolution.

Since we simulate the dictionary, the proposed Latent Signal Model framework does not require acquisition of any fully-sampled datasets for model training. In addition, the auto-encoder is trained once and can be reused with subsequent acquisitions from the same sequence.

Informed by Supporting Figures 1 and 2, we used the hyperbolic tangent nonlinearity with either two or three fully connected layers in the encoder and decoder. This extends well into EPTI and MPRAGE-shuffling, but ideal hyper-parameters may change in other applications.

Most of the reconstruction experiments in this work employ regularization in the proposed and linear reconstructions to make comparisons as fair and competitive as possible. We included Figure 5 to isolate the effects from reduced degrees of freedom and understand our technique's sensitivity to different noise instances. As the figure suggests, the reduced degrees of freedom provides reliable improvements in reconstruction quality. Combining the proposed

Latent Signal Model approach with spatial regularization then yields additive improvements, as demonstrated in the other experiments.

In the $T_2$-shuffling setting, Figure 6 demonstrated that the proposed technique finds local minima of reconstruction problems that incorporate EPG-simulations in the forward model. In comparison to solving the EPG-based reconstruction problem, we found that our proposed approach reduced reconstruction time per iteration by at least an order of magnitude. Additionally, when not initialized with the proposed framework's solution, we found that the EPG based reconstruction problem does not converge with the simulated data. Thus, we envision our approach being advantageous when differentiation through signal simulation requires significant computation. For example, applications that require iso-chromat based simulations[27,60] might benefit significantly from the proposed approach.

Figure 7 and Supporting Figure 3 compare reconstructions from highly-undersampled EPTI data with phase estimated from fully-sampled and low-resolution data respectively. The linear reconstruction breaks down in both regimes, while Latent Signal Models produce reasonable results using both phase estimates. However, the proposed approach improves significantly when utilizing the fully-sampled phase estimate, in comparison to the low-resolution estimate. Future work will explore improved $B_0$ estimation and refinement[47,61,62] techniques to improve phase-estimates for Latent Signal Model EPTI reconstructions.

*Limitations*

The presence of neural networks in the proposed framework induces a non-convex reconstruction problem. Thus, the optimization may converge to a suboptimal local minimum with an improperly trained auto-encoder. For example, the LeakyRelu autoencoders, explored in Supporting Figures 1 and 2, yield significant variability in performance from different random initialization of model weights. We found that tanh activations produce reconstruction problems that reliably terminate at reasonable local minima. Avoiding sub-optimal local minima in other applications may require changes to the auto-encoder structure.

In the linear forward model, from Eq1, the Fourier and coil operator, $FS$, commutes with the temporal operator, $\Phi^6$, when the phase operator is $H = 1$, significantly reducing computation in the shuffling regime. However, $FS$ does not commute with $Q$ in Eq3, so our technique requires increased computational complexity in comparison to linear shuffling. For example, in simulated and retrospective $T_2$-shuffling reconstructions, Latent Signal Models took 60 ms and 32 ms per iteration respectively when running on a GPU. BART linear shuffling reconstructions took 300 ms and 200 ms per iteration respectively but ran on a CPU. If run on a GPU, BART iterations would most likely be comparable or faster than Latent Signal Model iterations. Note, these operators do not commute when $H \neq 1$, so the proposed framework requires similar levels of computation to linear EPTI.

## 5   CONCLUSION

With the proposed Latent Signal Model framework, directly solving for learned nonlinear latent representations of signal improves time-resolved MRI reconstruction through reduced degrees of freedom. $T_2$-shuffling, EPTI, and MPRAGE-shuffling applications demonstrate quantitative and qualitative benefits of the proposed technique in simulation and in-vivo experiments.

## DATA AVAILABILITY STATEMENT

Code and data used to generate Figures 2 - 8 can be found in the following two links respectively:

https://github.com/YaminArefeen/latent_signal_models_mrm_2022

https://www.dropbox.com/sh/5upn7121m0dxvjp/AAD2liymnq4LhNKs85nUddJya?dl=0

## ACKNOWLEDGEMENTS

The authors thank Siddharth Iyer for his helpful advice and the MPRAGE-Shuffling data. This work was supported in part by research grants NIH R01 EB017337, U01 HD087211, R01HD100009, R01 EB028797, U01 EB025162, P41 EB030006, U01 EB026996, R03EB031175 and the NVidia Corporation for computing support. This material is also based upon work

supported by the National Science Foundation Graduate Research Fellowship Program under Grant No. 1122374. Any opinions, findings, and conclusions or recommendations expressed in this material are those of the authors and do not necessarily reflect the views of the National Science Foundation. In addition, research reported in this manuscript was supported by the National Institute of Biomedical Imaging and Bioengineering (NIBIB), of the National Institutes of Health under award number 5T32EB1680.

# Supporting Figure 1

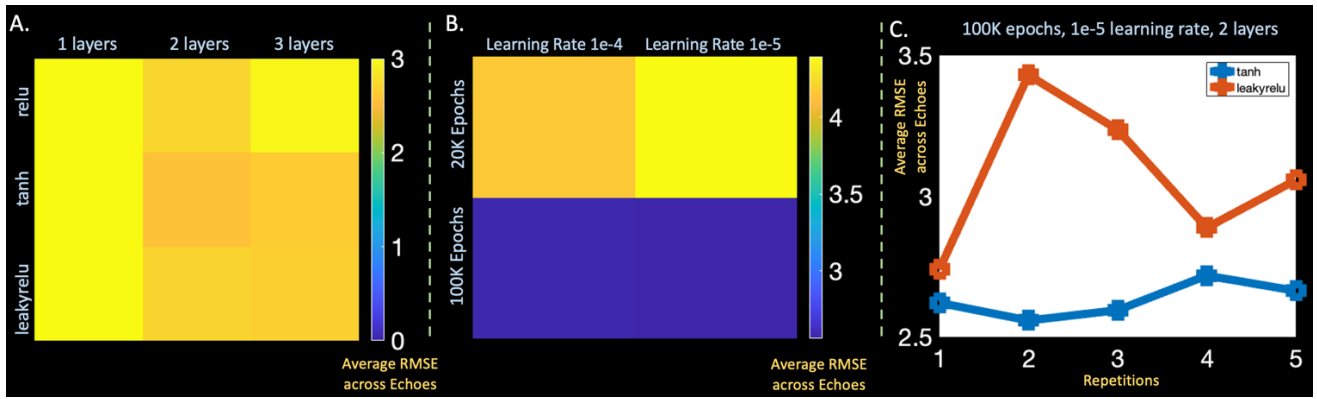

**Supporting Figure 1 Caption:** Ablation experiment comparing the performance of different auto-encoder model hyper-parameters in simulated reconstructions. (A) Grid of average RMSE across all echoes comparing models trained with different non-linearities and number of layers. Hyperbolic tangent (tanh) with 2 layers yields the best results. (B) Average RMSE for tanh with 2 layers using different learning rates and step-sizes. 100K epochs with 1e-5 learning rate gives the lowest RMSE. (C) Reconstruction RMSE of tanh and leakyrelu (both with 100K epochs, 2 model layers, 1e-5 learning) for various random initialization of model parameters. LeakyRelu varies significantly with random initialization while tanh yields more consistent results.

# Supporting Figure 2

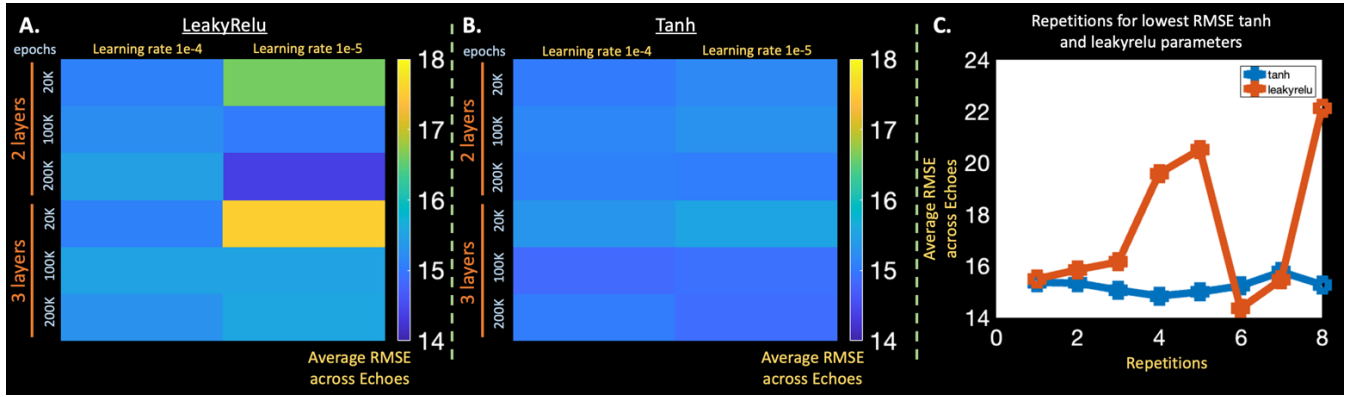

**Supporting Figure 2 Caption:** Ablation experiment comparing the performance of different auto-encoder model hyper-parameters in in-vivo retrospective reconstructions. (A) and (B) display grids of average RMSE across all echoes for models with LeakyRelu and Tanh for a range of layers, learning rates, and epochs. LeakyRelu achieves lowest RMSE with 2 layers, 200K epochs, and 1e-5 learning rate, while Hyperbolic tangent achieves its minimum with 3 layers, 100K epochs, and 1e-4 learning rate. (C) Plots the performance of LeakyRelu and Tanh, with their respective best hyper-parameters, at 8 different random initializations. LeakyRelu varies significantly, while tanh yields consistent results.

# Supporting Figure 3

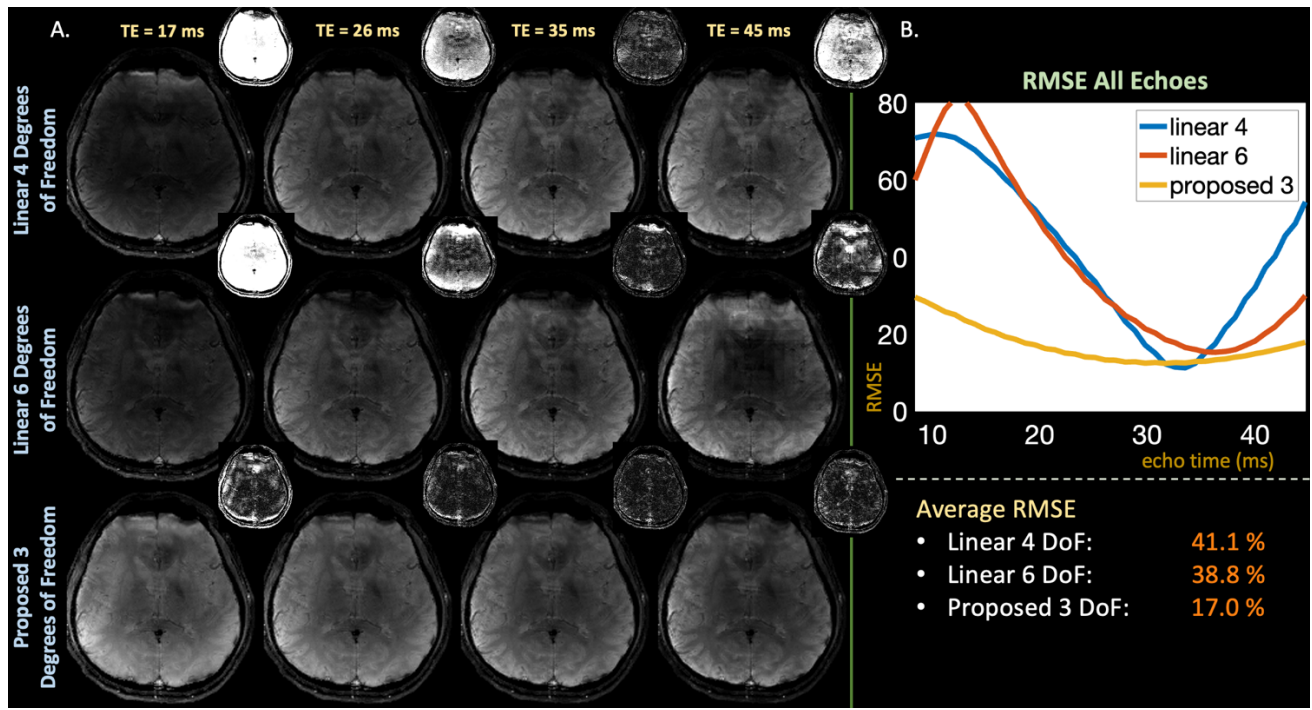

**Supporting Figure 3 Caption:** Similar comparisons of the proposed and linear subspace approach from Figure 7 using phase estimated from low-resolution calibration data in the GE-EPTI forward model. (A) displays exemplar echo images and associated error maps and (B) plots NRMSE for each echo. Even with a worse phase-estimate, the proposed approach still produces significantly higher quality images with lower NRMSE in comparison to the linear techniques.

# FIGURE CAPTIONS

**Figure 1:** Comparing linear subspace constraints and the proposed Latent Signal Model framework in the $T_2$-shuffling setting. To represent FSE signal evolution, (A) the SVD generates a subspace from a dictionary of signal evolution with 2-4 coefficients, while (B) the auto-encoder learns a more compact representation of signal evolution with 1 latent variable. Inserting either the subspace or decoder in the forward model reduces the number of variables required to resolve signal dynamics. The auto-encoder's more compact latent representation, compared to linear subspaces', enables improved reconstruction quality in subsequent experiments.

**Figure 2:** (A) + (B) Average error across the entire training and testing dictionaries of FSE signal evolution using linear subspaces and proposed auto-encoders with {1,2,3,4} coefficients or latent variables. (c) Error for individual signal evolutions, associated with different T2 values, in the testing dictionary using subspaces and auto-encoders. The subspace requires 3-4 coefficients, while the auto-encoder effectively captures signal evolution with just 1 latent variable.

**Figure 3:** (A) Exemplar reconstructed echo-images, associated error maps, and (B) RMSE plot vs. echo time from a simulated T2-shuffling acquisition comparing the subspace and proposed Latent Signal Model reconstruction with {4,6} and 3 degrees of freedom respectively. The reduced degrees of freedom enabled by inserting the decoder into the $T_2$-shuffling forward model yields cleaner images and lower RMSE.

**Figure 4:** (A) Exemplar reconstructed echo images and associated error maps from the in-vivo, 7-shot retrospective, 2D-$T_2$-shuffling acquisition at exemplar echo times comparing subspace constraints with {4,6} and the proposed approach with 3 degrees of freedom. (B) RMSE plots comparing reconstructed images at all echo times. Like in simulation, the proposed approach yields cleaner images and lower RMSE through reduced degrees of freedom.

**Figure 5:** (A, C) Exemplar average error maps for the proposed and linear reconstructions across the 250 k-space instances in both the simulated and retrospective datasets. (B,D) Average and standard deviation of reconstruction RMSE at each echo. With its reduced degrees of freedom, the proposed Latent Signal Model approach achieves lower average error while maintaining similar variance across the k-space instances in-comparison to reconstructions with linear subspace constraints.

**Figure 6:** (B, D) Gradient norms with respect to $T_2$ and density of the simulation based forward model as a function of the Latent Signal Model optimization iteration. (A,C) exemplar reconstructed echo images from the proposed approach and the EPG-based forward model initialized with the proposed approach. The gradient norms approaching zero and the similarity in the reconstructions suggest that the proposed approach effectively and efficiently finds a solution to the EPG-based forward model.

**Figure 7:** (A) Exemplar reconstructed echo images and associated error maps from the GE-EPTI dataset using phase estimated from the fully-sampled dataset, and (B) plots comparing NRMSE at each individual echo image. The proposed Latent Signal Model approach with reduced degrees of freedom significantly improves qualitative image quality and quantitative NRMSE in comparison to the linear subspace constrained reconstructions.

**Figure 8:** (A) + (B) Average error across the entire training and testing dictionaries of MPRAGE signal evolution using linear subspaces and proposed auto-encoders with {1,2,3,4} coefficients or latent variables. (c) Error for individual signal evolutions, associated with different $T_1$ values, in the testing dictionary using subspaces and auto-encoders. The subspace requires 3-4 coefficients, while the auto-encoder effectively captures signal evolution with just 1 latent variable.

**Figure 9:** Reconstructions with and without regularization on the MPRAGE shuffling dataset comparing the proposed Latent Signal Model framework and linear subspace constraints. Without regularization, the proposed approach significantly reduces noise amplification and improves reconstruction quality. With applied regularization, linear + wavelet suffers increased blurring while linear + locally-low-rank still exhibits noise amplification. The proposed approach reduces reconstruction artifacts, particularly at TI ≈ 500 ms, improves image sharpness in comparison to linear + wavelet, and reduces noise amplification in comparison to linear + locally-low-rank.

## Supporting Figure Captions

**Supporting Figure 1:** Ablation experiment comparing the performance of different auto-encoder model hyper-parameters in simulated reconstructions. (A) Grid of average RMSE across all echoes comparing models trained with different non-linearities and number of layers. Hyperbolic tangent (tanh) with 2 layers yields the best results. (B) Average RMSE for tanh with 2 layers using different learning rates and step-sizes. 100K epochs with 1e-5 learning rate gives the lowest RMSE. (C) Reconstruction RMSE of tanh and leakyrelu (both with 100K epochs, 2 model layers, 1e-5 learning) for various random initialization of model parameters. LeakyRelu varies significantly with random initialization while tanh yields more consistent results.

**Supporting Figure 2:** Ablation experiment comparing the performance of different auto-encoder model hyper-parameters in in-vivo retrospective reconstructions. (A) and (B) display grids of average RMSE across all echoes for models with LeakyRelu and Tanh for a range of layers, learning rates, and epochs. LeakyRelu achieves lowest RMSE with 2 layers, 200K epochs, and 1e-5 learning rate, while Hyperbolic tangent achieves its minimum with 3 layers, 100K epochs, and 1e-4 learning rate. (C) Plots the performance of LeakyRelu and Tanh, with their respective best hyper-parameters, at 8 different random initializations. LeakyRelu varies significantly, while tanh yields consistent results.

**Supporting Figure 3:** Similar comparisons of the proposed and linear subspace approach from Figure 7 using phase estimated from low-resolution calibration data in the GE-EPTI forward model. (A) displays exemplar echo images and associated error maps and (B) plots NRMSE for each echo. Even with a worse phase-estimate, the proposed approach still produces significantly higher quality images with lower NRMSE in comparison to the linear techniques.